\documentstyle[aps,eqsecnum]{revtex}
\input BoxedEPS.tex
\tighten

\title{Short-range interactions in an effective field 
theory approach for nucleon-nucleon scattering}

\author{K.~A. Scaldeferri, D.~R. Phillips, C.-W. Kao, T.~D. Cohen}

\address{Department of Physics, University of Maryland, College Park,
MD, 20742, USA}

\begin{document}
\SetRokickiEPSFSpecial

\maketitle

\date{\today}
\begin{abstract}
We investigate in detail the effect of making the range of the ``contact''
interaction used in effective field theory (EFT) calculations of NN
scattering finite. This is done in both an effective field
theory with explicit pions, and one where the pions have been
integrated out. In both cases we calculate NN scattering in the
${}^1 S_0$ channel using potentials which are second-order in the EFT
expansion. The contact interactions present in the EFT Lagrangian are
made finite by use of a square-well regulator. We find that there is
an optimal radius for this regulator, at which second-order
corrections to the EFT are identically zero; for radii near optimal
these second-order corrections are small. The cutoff EFTs which
result from this procedure appear to be valid for momenta up to about
100 MeV/c. We also find that the radius of the square well cannot be
reduced to zero if the theory is to reproduce both the experimental
scattering length and effective range. Indeed, we show that, if the NN
potential is the sum of a one-pion exchange piece and a short-range
interaction, then the short-range piece must extend out beyond 1.1 fm,
regardless of its particular form.
\end{abstract}

\section{Introduction}

Effective field theory (EFT) techniques have been used successfully
for many years to study problems in particle physics where a
well-defined hierarchy of mass scales exists~\cite{KM95}.  
In such problems one can, in principle,
integrate out the short-range degrees of freedom, {\it i.e.} those
corresponding to momenta above some 
separation scale, thereby obtaining a non-local effective action which will
be applicable at energies well below this scale.
The assumption is then made that all 
momenta in the problem of interest are small compared to the 
masses of the degrees of freedom which were integrated out.
It follows that a momentum expansion of the action may be made. This leads
to calculations which are organized in terms of the power of 
momentum which a given diagram contributes---the so-called EFT expansion.

This is possible because power-counting arguments indicate that a given loop 
diagram only contributes at a definite order in the EFT expansion. Naturally,
these loop diagrams are divergent, and must be 
regulated if they are to yield finite results. 
These results must then be renormalized.
The theory as a whole is not renormalizable, but at a given order in the EFT
expansion there are only a finite number of possible counterterms and so
the theory does have predictive power. Hence, once
the order of an EFT calculation is fixed, the predictions of that theory are
independent of the regulation scheme chosen.

Several years ago, Weinberg suggested that the techniques of EFT could
be modified to study low-energy problems in nuclear physics, including
nucleon-nucleon scattering~\cite{We90,We91}. He pointed out that the
fundamental difference in the NN scattering case is that the EFT
expansion must be applied to the potential---not the amplitude, as in
most other problems.  If this is done then the potential which is to
be used in the Schr\"odinger equation contains a delta function
interaction.  Direct calculation with such an equation is impossible.
Regulation and renormalization of the results must be performed before
physical predictions can be extracted. The question is how this is to
be done.  

So far two approaches exist in the literature. The first involves
formally iterating the divergent interaction, and then renormalizing
the resultant amplitude. This was the approach used by
Weinberg~\cite{We91}. More recently, it has been applied by Kaplan
{\it et al.}, who iterated the contact interaction formally and then
used dimensional regularization and the $\overline{{\rm MS}}$
renormalization scheme to calculate NN scattering in the ${}^1S_0$
channel~\cite{Ka96}. The second approach is to introduce a regulator
into the equation before iterating the potential to generate the
t-matrix.  This was the method adopted by van Kolck {\it et al.}, who
cut off all integrals in momentum space in studies of NN
scattering~\cite{vK96}.  In such an approach renormalization is
performed by adjusting the coefficients of the potential to fit the NN
scattering data. This can be done for several different values of the
regulator parameter. The sensitivity of unfit physical observables to
the regulator parameter may be regarded as a measure of the validity
of the regulation.  If an EFT solution to the NN problem can be
constructed by one of these two, or some other, means, it may then be
used as input in nuclear physics applications of EFT, e.g. to
pion-deuteron scattering~\cite{We92}, the three-nucleon
system~\cite{vK94}, the $pp \rightarrow pp\pi^0$
process~\cite{vK95,Ku95}, and pion photoproduction on the
deuteron~\cite{Be95}.

One might think that in order to establish the connection with the
original field theory the regulator in the second approach must
ultimately be removed from the problem, {\it i.e.}, that the range
of the regulated contact interaction must eventually be taken to zero.
However, it has recently been pointed out that the use of zero-range
interactions which are defined as the limits of short-range ones as
the range is taken to zero leads to peculiar results. In particular it
has been shown that EFTs either with or without pions cannot contain a
repulsive zero-range interaction (defined in this sense) which has any
effect on the observables in the renormalized theory~\cite{Co96,CP96}.
Thus all non-trivial zero-range interactions must be attractive.
Furthermore, in theories without explicit pions ({\it i.e.} where the
pions are integrated out) it was shown that if the range
of the interaction is taken to zero then the phase shifts obey:
\begin{equation}
\frac{d}{d k^2} (k \cot(\delta(k))) \leq 0.
\end{equation}
A corollary is that potentials of sufficiently small radii {\em
cannot} reproduce the NN ${}^1S_0$ scattering length and effective
range. In this paper we show that similar constraints apply when pionic
degrees of freedom are explicitly included.

While these results do not invalidate the conclusions of
Ref.~\cite{Ka96} they do show that the approach used there is {\it
not} equivalent to solving a Schr\"odinger equation with an
interaction of range $R$, and renormalizing the coefficients as $R$ is
adjusted. In other words, for this problem, regulating the potential
and then iterating it is not equivalent to iterating it formally and
then using dimensional regularization and $\overline{\rm{MS}}$
renormalization.

This leads us to investigate in detail what happens if the range of
the ``contact'' interaction is made finite before iteration, as was
done in practice in Ref.~\cite{vK96}. (The use of similar ``cutoff''
field theories has been advocated for some time by, for example,
Lepage~\cite{Le89}.)  While such an approach apparently violates some
of the assumptions of Weinberg's power-counting arguments~\cite{We91},
one trusts that for a wide range of regulator parameter values the
results of the renormalized theory are not particularly sensitive to
the exact value of that parameter.  Indeed, a virtue may be made of
the necessity of keeping this parameter finite. The freedom to choose
it may be exploited so as to minimize the error in truncating the EFT
expansion.  That is, a range may be chosen for our short-range
interaction which results in the renormalized coefficient of the
second-order term in the EFT expansion being zero. For other values of
the short-range interaction range the value of this renormalized
second-order coefficient will, of course, be non-zero.  The viability
of such an approach must be judged by testing how sensitive
observables are to the range chosen for the regulated contact
interaction.

In this paper we explore this issue for two EFT-motivated treatments
of nucleon-nucleon scattering in the $^1S_0$ channel~\cite{RvK96}. In
Section~\ref{sec:sqwell} we investigate a Lagrangian with only contact
interactions ({\it i.e.} where the pions have been integrated out),
while in Section~\ref{sec:pion} we look at an EFT with contact
interactions and explicit pions. In both cases we never use a true
contact interaction but instead always solve the Schr\"odinger
equation keeping the regulator parameter for the contact interaction,
$R$, finite. The particular form of the regulator chosen here is a
square well.  In both EFT treatments we begin by calculating at zeroth
order in the EFT expansion. At this order, the potential in the
Schr\"odinger equation is either a square well alone, or the sum of a
square well and one-pion exchange.  For any given radius $R$ the
strength of the well can be adjusted to match the $\mbox{}^1 S_0$ NN
scattering length.  However, by exploiting the freedom to choose $R$,
we can minimize the second-order corrections to the EFT by choosing a
radius which results in the total zeroth-order EFT potential
reproducing {\it both} the scattering length and the effective range.
The resulting value of the well radius, $R$, is the ``optimal'' one,
since it minimizes the second-order EFT corrections to the potential.
For an arbitrary regulator radius $R$, going to second order in the
EFT means adding terms to the potential which correspond to
derivatives of our ``contact'' interaction.  The coefficients of these
new terms are fixed by demanding that at different well radii the
scattering length and effective range are still reproduced. At the
optimal radius these coefficients are, by definition, zero. For radii
close to optimal, it is found that the coefficients can be
renormalized successfully.  However, we will show that there is a
lower bound on the radius for which this renormalization can be done.
Below a certain value of $R$ it is impossible to adjust the
coefficients to reproduce low-energy ${}^1S_0$ NN scattering data.
This explicitly demonstrates, for the specific case of a square-well
regulator, that the theorem of Ref.~\cite{CP96} prevents one from
taking the limit $R \rightarrow 0$. More generally, the result of
Ref.~\cite{CP96} leads to an absolute lower bound on the radius $R$
that may be used if the effective field theory is to correctly predict
the effective range. This lower bound is completely independent of the
particular regulator chosen.  In the EFT without explicit pions it is
a simple matter to show that the absolute lower bound is $R=1.3$
fm. In Section~\ref{sec-lowerbd} we use a modified version of the
argument presented in Ref.~\cite{CP96} to show that the smallest
possible $R$, even when explicit one-pion exchange is included, is 
larger than 1.1 fm.

The sensitivity of the results to the range of the short-range potential 
is explored in two ways. First, we examine the
phase shifts for different values of $R$.  Agreement is expected at low
momenta since we have fit the scattering length and effective range.
Thus the phase shifts should be identical until the point where the
fourth-order terms in the effective range expansion
become significant.  The question is at what momentum 
these fourth (and higher) order terms become important for different well 
radii. Secondly, we examine the magnitude of the second-order
terms in the t-matrix for potentials of various radii. The hope is that
for radii close to optimal the second-order corrections are
small for moderate momenta. This would indicate that if short-range
potentials of these radii are used then the EFT expansion stays under control.

\section{An Effective Field Theory Without Pions:
The Square Well and Derivatives}
\label{sec:sqwell}

As observed in Refs.~\cite{We90,Ka96}, if one integrates out all the exchanged 
mesons in the NN interaction then an EFT is obtained which consists solely
of contact interactions and derivatives thereof. In the ${}^1 S_0$ channel
the only pieces of this Lagrangian which contribute (to second order 
in the EFT expansion) are:
\begin{equation}
{\cal L}=N^\dagger i \partial_t N - N^\dagger \frac{\nabla^2}{2 M} N
- \frac{1}{2} C_S (N^\dagger N)^2 - \frac{1}{2} C_T (N^\dagger {\bf \sigma}
N)^2 - \frac{1}{4} C_2 (N^\dagger \nabla^2 N) (N^\dagger N) -
\frac{1}{4}  C_2 ((\nabla^2 N^\dagger) N) (N^\dagger N).
\label{eq:L}
\end{equation}

If the interaction terms here are used, as advocated by Weinberg, to generate
a potential for use in the Schr\"odinger equation, then at second order in 
the EFT expansion the result is a
Schr\"{o}dinger equation which we will write with a general 
non-local potential,
\begin{equation}
-\frac{\nabla^2}{M} \psi({\bf x}) + \int V_R({\bf x,x'})
\psi({\bf x'}) d^3 x' = E \psi({\bf x}),
\label{eq:Sch}
\end{equation}
where the subscript $R$ indicates that $V$ must be regulated
in order to make the Schr\"odinger equation meaningful.
The regulated non-local potential is:
\begin{equation}
V_R({\bf x},{\bf x'}) = (C_0 + C_2 ( - \nabla^2 - \nabla'^2))
\delta^{(3)}({\bf x} - {\bf x'}) \delta_R^{(3)}({\bf x}).
\label{eq:VR}
\end{equation}
Here $C_0=C_S - 3C_T$, and $\delta_R^{(3)}$ is a function which tends
to a delta function as $R \rightarrow 0$.  In this work we use a
square well of width $R$ for the function $\delta_R^{(3)}$; {\it
i.e.},
\begin{equation}
\delta_R^{(3)}({\bf x}) = \frac{3 \Theta(R-|{\bf x}|)}{4 \pi R^3}.
\label{eq:sqwell}
\end{equation}

In the $l = 0$ partial wave, the radial equation obtained from the
Schr\"odinger equation (\ref{eq:Sch}) is then
\begin{equation}
\left( -\frac{1}{M} - 2 C_2 \delta_{R}(r) \right) \frac{d^2
U}{dr^2} - 2 C_2 \left( \frac{dU}{dr} - \frac{U(r)}{r} \right)
\frac{d \delta_R}{dr} + \left( C_0 \delta_{R}(r) - C_2 \frac{d^2
\delta_{R}}{dr^2} \right) U(r) = E U(r).
\label{eq:radial}
\end{equation}
The solutions to this equation in the two regions $r < R$ and $r > R$
must be matched at $r=R$. By integrating Eq.~(\ref{eq:radial}), we see
that the boundary condition at $r=R$ involves a discontinuity  
of the derivative of the wavefunction, due to the derivative terms in
the potential:
\begin{equation}
\left( 1 + \frac{3 C_2 M}{4 \pi R^3} \right) \left. \frac{dU}{dr}
\right|_{-} = \left. \frac{dU}{dr} \right|_{+},
\label{eq:bound}
\end{equation}
where $dU/dr|_{\pm}$ denote the one-sided derivatives at $r = R$. 

From Eq.~(\ref{eq:bound}) an expression for the phase-shift $\delta$
at momentum $k = ( M E )^{1/2}$ may be derived.  It is:
\begin{equation}
k \cot \delta = \frac{B \kappa \cot(\kappa R) + k \tan(k R)}
	{1 - B \frac{\kappa}{k} \cot(\kappa R) \tan(k R)},
\label{kcotd}
\end{equation}
where
\begin{equation}
B = 1 + \frac{3 C_2 M}{4 \pi R^3}, \qquad 
\kappa = \left( \frac{k^2 - \frac{3 C_0 M}{4 \pi R^3}}
	{1 + \frac{3 C_2 M}{2 \pi R^3}} \right)^{1/2}.
\end{equation}

Expanding Eq.~(\ref{kcotd}) in powers of $k^2$, we find that the
scattering length, $a$, and effective range, $r_e$, of this potential
are given by
\begin{equation}
a = R - \frac{R}{B x \cot x},
\label{eq:a}
\end{equation}
\begin{equation}
r_e = \frac{R(R - a)}{a} \left[ \frac{2 \pi R^4 ( 1 - x \tan x - 
	x \cot x)}{(2 \pi R^3 + 6 C_2 \mu) x^2 a} - 2
	\right],
\label{eq:re}
\end{equation}
where
\begin{equation}
x = R \left( \frac{ - 3 C_0 M}{4 \pi R^3 + 6 C_2 M} \right)^{1/2}.
\label{eq:x}
\end{equation}

As we can see from Eqs.~(\ref{eq:a}) and (\ref{eq:re}) above, $x$
is a more natural variable for this problem than $C_0$.  Consequently,
from now on we will generally use $x$ and $C_2$ as variables.

The values adopted for the NN ${}^1S_0$ scattering length and effective range
here are:
\begin{equation}
a=-24 \mbox{ fm}; \qquad r_e=2.7 \mbox{ fm}.
\label{eq:exptalvals}
\end{equation}
We will examine how $x$ and $C_2$ must be renormalized as $R$ is
varied in order to reproduce these values.  We will begin by setting
$C_2 = 0$ and solving Eqs.~(\ref{eq:a}) and (\ref{eq:re}) numerically
for $R$ and $x$.  We will refer to the result as the ``optimal'' well
since it is the well which minimizes the second-order corrections in
the EFT expansion.  We will then vary $R$ from the optimal value and
solve for $x$ and $C_2$ to determine over what range of $R$ we can
successfully obtain the actual scattering length and effective range
using this potential. Finally, we will examine the phase shifts and
the second-order corrections to the t-matrices for these renormalized
potentials and compare our results with the phase shifts extracted
from the experimental data.

Before we begin this procedure note that Phillips and Cohen
\cite{CP96} have recently shown that a bound for $d\delta/dk$ derived
by Wigner~\cite{Wi55} yields the following constraint on $r_e$:
\begin{equation}
r_e \leq 2 \left( R - \frac{R^2}{a} + \frac{R^3}{3 a^2} \right).
\label{eq:rmax}
\end{equation}
Eq.~(\ref{eq:rmax}) provides an absolute lower bound on the size of a
potential which can reproduce a given scattering length and effective
range.  In the $\mbox{}^{1}S_0$ channel for NN scattering, this lower
bound is $1.3$ fm.  In fact, for this potential, we will find that the
lowest value of $R$ for which Eqs.~(\ref{eq:a}) and (\ref{eq:re}) can
be solved is considerably larger than this.

\begin{figure}
\HideDisplacementBoxes
\centerline{\BoxedEPSF{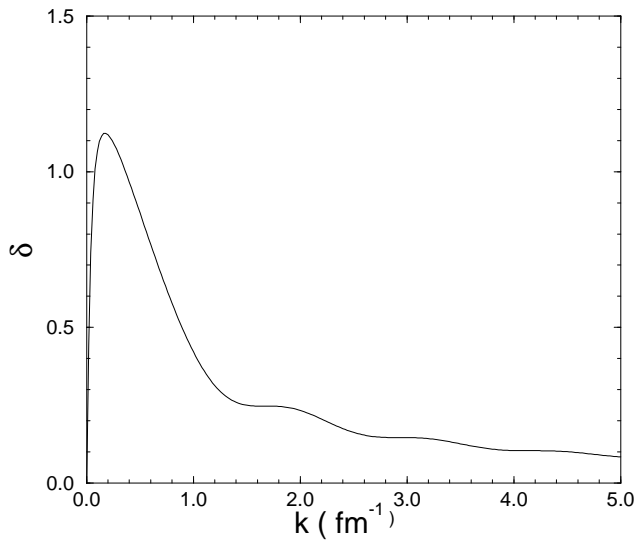 scaled 500}}
\caption{A plot of phase shifts in radians versus momentum for the
	optimal well. The phase shifts are calculated from
	Eq.~(\ref{kcotd}) with $R = 2.6\ {\rm fm}, x =
	1.5$ and $C_2 = 0$.}	\label{f:r0} 
\end{figure}

To determine the optimal width of the
square well, Eqs.~(\ref{eq:a}) and (\ref{eq:re}) are solved with
$C_2 = 0$, with the result that the optimal width is $2.6$ fm. The
corresponding value of $x$ is $1.5$. These values of $x$, $C_2$, and $R$ are
equivalent to $C_0 =-5.1\ {\rm fm}^2$. A plot of the phase shift as a
function of $k$ for these values of $R, x$ and $C_2$ is shown in
Figure~\ref{f:r0}. We confirm that, as we expect from Levinson's
Theorem for a potential with no bound states, the
phase shifts at zero and very large energy are equal.

\begin{figure}
\HideDisplacementBoxes
\centerline{\BoxedEPSF{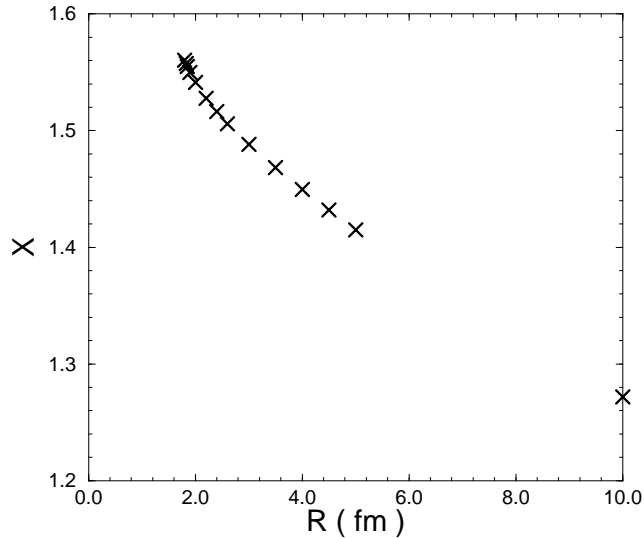 scaled 500}}
\caption{A plot of the renormalized values of $x$, defined as in
	Eq.~(\ref{eq:x}), versus the radius, $R$, of a square well potential.}
		\label{f:x}
\end{figure}

\begin{figure}
\HideDisplacementBoxes
\centerline{\BoxedEPSF{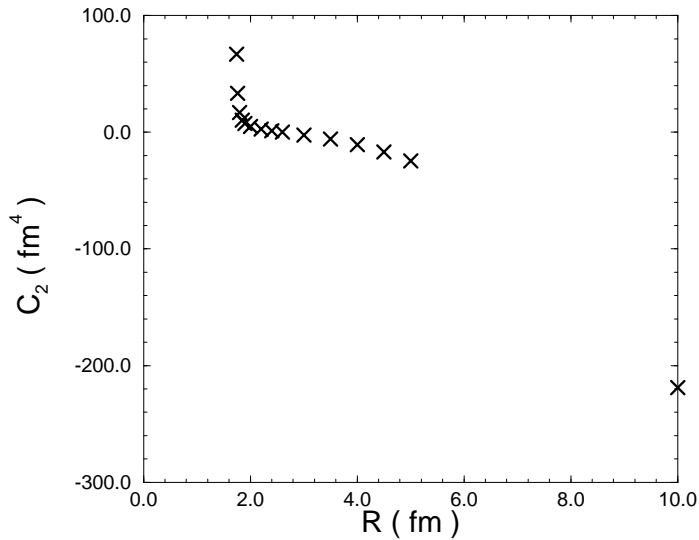 scaled 500}}
\caption{A plot of the renormalized values of the second-order
	coefficient in the EFT expansion, $C_2$, versus the radius,
	$R$, of a square well potential.  The value of $C_2$ diverges
	near $R = 1.7$ fm.}
		\label{f:c2}
\end{figure}

We now proceed to examine the behavior for other values of $R$.  When
Eqs.~(\ref{eq:a}) and (\ref{eq:re}) are solved numerically for $x$ and
$C_2$ as $R$ is varied from the optimal value, we find the behavior
shown in Figures~\ref{f:x} and \ref{f:c2}.  As the radius is reduced,
we find that $x$ tends to $\pi / 2$ and $C_2$ diverges as we approach
a radius near $1.7\ \rm{fm}$.  The fact that $x$ is tending to a
finite quantity while $C_2$ diverges suggests that $C_0$ is also
diverging.  For large $R$, $C_2$ becomes large and negative.

\begin{figure}
\HideDisplacementBoxes
\centerline{\BoxedEPSF{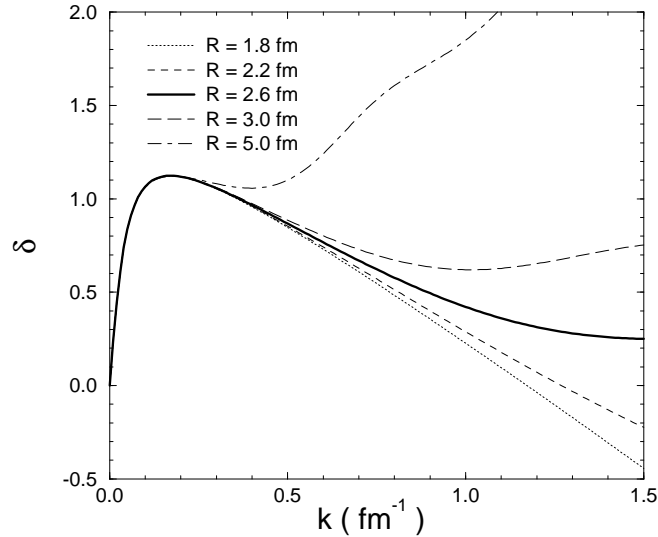 scaled 500}}
\caption{A plot of the phase shifts in radians versus momentum
	calculated from Eq.~(\ref{kcotd}) for square wells with
	different radii.  The phase shifts for the optimal well, shown
	in more detail in Figure~\ref{f:r0}, are shown here in bold
	for comparison.}  
		\label{f:rall} 
\end{figure}

Armed with renormalized values of $x$ and $C_2$ for several different
radii, we can use Eq.~(\ref{kcotd}) to examine how the phase shifts
are affected by changing $R$.  Figure~\ref{f:rall} compares the phase
shifts for several values of $R$ with those obtained at the optimal
value of $R$.  Since we have fit the scattering length and effective
range, the phase shifts will agree for all radii until the fourth
order terms in the effective range expansion become important. While
this occurs at quite low momentum for the 5 fm well, when the radii
are close to optimal the agreement persists to a much higher momentum,
$k \approx 0.5 \mbox{ fm}^{-1}$.  This is roughly what we expect since
the radii of the relevant wells are themselves around 2 fm. Since this
is a theory without explicit pions this radius can be thought of as
being intrinsically that of one-pion exchange.  Thus one might hope
that when pions are explicitly included in the theory the radius of
the short-range potential will decrease significantly.

For non-optimal wells the phase shifts always go to positive or
negative infinity as the momentum gets large.  However, this does not
contradict Levinson's Theorem since these potentials are non-local.

\begin{figure}
\HideDisplacementBoxes
\centerline{\BoxedEPSF{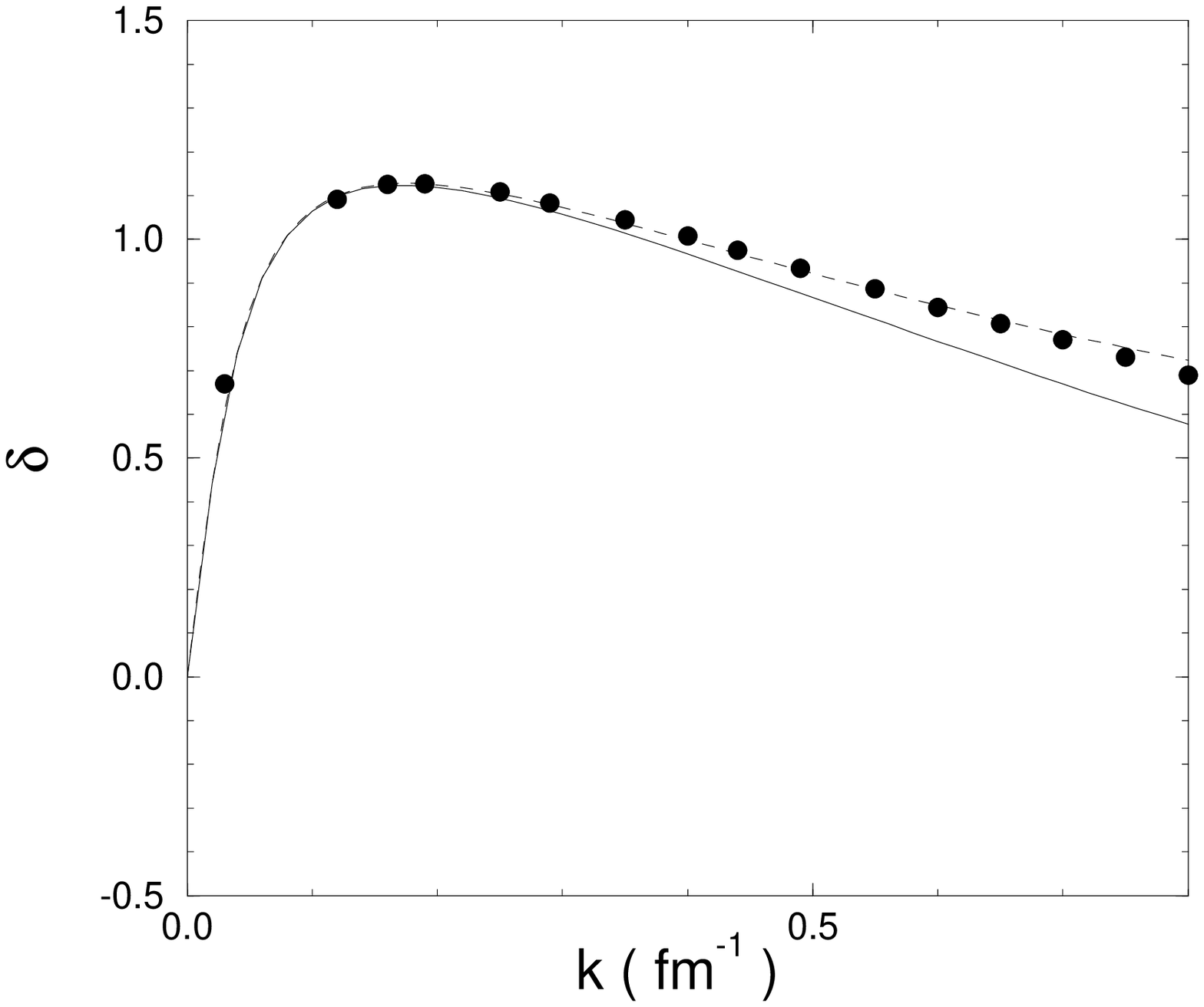 scaled 500}}
\caption{A comparison of calculated and experimentally measured phase
	shifts. The solid line in this plot is the phase shift as a
	function of momentum for an optimal square well.  The dashed
	line is the effective range expansion to second order.  The
	dots are data points from the Nijmegen $np$ partial wave
	analysis in the $\mbox{}^1 S_0$ channel~\protect\cite{St93}.}
		\label{f:nijm} 
\end{figure}

Next we compare these phase shifts with the experimentally determined
values.  Figure~\ref{f:nijm} compares the data from the Nijmegen
partial wave analysis~\cite{St93} with the phase shifts from the
optimal well and the phase shifts from the effective range expansion.
Unfortunately, we find rather poor agreement with the data for our
optimal square well.  Contrastingly, the effective range expansion
matches the data surprisingly well, indicating that the shape
parameter for NN scattering in this channel is quite small.  Of
course, we could adjust all three parameters $R$, $C_0$, and $C_2$ in
order to improve the agreement of our result with the data, but this
would not be in the spirit of the calculations we are pursuing here.
While the result displayed in Fig.~\ref{f:nijm} is not particularly
promising for the approach espoused in this paper, we will see that
the explicit inclusion of pion exchange will lead to a considerable
improvement in the fit to the experimental data. 

Having examined the behavior of this system to second order in the EFT
expansion, we now wish to examine just how significant
the second order, {\it i.e.} derivative, terms are.  For a given value of $R$,
we may examine the fractional difference between the on-shell
t-matrices at zeroth and second order in the EFT expansion, 
\begin{equation}
\Delta T_{R}(k) \equiv \frac{T^{(0)}_{R}(k) - T^{(2)}_{R}(k)}{T^{(0)}_{R}(k)}.
\label{eq:DeltaT}
\end{equation}
Here, the second-order t-matrix, $T^{(2)}_{R}$ has already been calculated:
\begin{equation}
T^{(2)}_{R}(k) = \frac{2 \pi / M}{k \cot \delta - i k},
\end{equation}
where $k \cot \delta$ is given in Eq.~(\ref{kcotd}).
On the other hand, $T^{(0)}_{R}$ is the zeroth-order EFT amplitude for a
square well of radius $R$.  Hence, at a given $R$ we have $C_2 = 0$
and the only free parameter is $C_0$. Consequently, we must
renormalize differently. In this case we use Eq.~(\ref{eq:a}) to fix
$x$ by demanding that we reproduce the $\mbox{}^{1}S_0$ NN scattering
length. This gives the renormalization condition
\begin{equation}
x_0 \cot x_0 = \frac{R}{R-a}.
\end{equation}
$T^{(0)}_{R}$ is then found by using the values $x = x_0$ and $C_2 = 0$ in
Eq.~(\ref{kcotd}) for $k \cot \delta$.

\begin{figure}
\HideDisplacementBoxes
\centerline{\BoxedEPSF{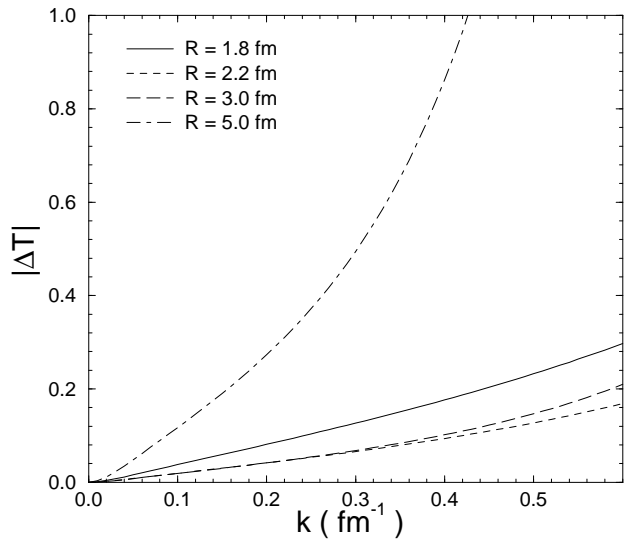 scaled 500}}
\caption{A plot of the fractional difference between the zeroth- and
	second-order t-matrices versus momentum, calculated from
	Eq.~(\ref{eq:DeltaT}), for several values of $R$, the radius of
	the square well.  For the optimal well with $R = 2.6$ fm, this
	quantity is identically zero.} 	\label{Tall}
\end{figure}

For the optimal width well, $\Delta T$ is zero, by construction.  For
non-optimal widths, Figure~\ref{Tall} shows $\Delta T_{R}(k)$ for
several values of $R$.  We note that, as we would expect, in the
zero-energy limit the difference is always zero.  The rate at which
$\Delta T_R$ grows with energy is dependent on how much the
radius of the well differs from that of the optimal well.

To summarize, we have shown how the coefficients in the second-order
EFT expansion must be renormalized if we wish to reproduce the
$\mbox{}^{1}S_0$ NN scattering length and effective range with a
square well potential of a given radius.  In particular, we have
demonstrated that there is a definite lower bound on the radius for
which this renormalization can be done.  For low momenta, we find that
the phase shifts are insensitive to the choice of regulator parameter,
as one would expect since we have fit the effective range expansion up
to second order. For radii close to the optimal value, we observe that
the agreement persists to higher momenta.  Furthermore, for these
nearly-optimal radii the quantity $\Delta T$ is small for momenta up
to roughly 0.5 fm$^{-1}$, thus indicating that the EFT expansion is
under control for momenta below this scale. Again, we would not expect
our EFT to be valid above this scale since it does not explicitly 
contain pions.

We do find, somewhat disappointingly, that a straightforward effective
range expansion of the data fits the actual phase shifts to
considerably higher momenta than our optimal well. This indicates that
the shape parameter for our optimal square well is considerably larger
than the experimental shape parameter, which appears to be very small.
While this situation could be remedied by adjusting our radius to
obtain a better agreement with the data such an approach would not be
systematic, as it would rely on the sensitivity of the phase shifts to
the cutoff parameter $R$.

\section{An effective field theory with explicit pions: 
One-Pion Exchange and a Square Well}

\label{sec:pion}

Naturally, the pionless EFT of Section~\ref{sec:sqwell} is only valid
for momenta considerably less than $m_\pi$.  In order to construct a
better EFT and so draw nearer to cases of interest in nuclear physics
in this section we include pions in our EFT.

At zeroth order the $NN$ potential resulting from such an EFT Lagrangian is:
\begin{equation}
V_R({\bf x}) = C_0 \delta_R^{(3)}({\bf x}) + V_\pi ({\bf x}),
\label{eq:EFT0}
\end{equation}
where $V_\pi({\bf x})$ is the ${}^1S_0$ OPE potential, which, if
we absorb the delta function piece into the contact interaction
takes the form
\begin{equation}
V_{\pi}(r) = - \alpha_{\pi} \frac{e^{-m_{\pi} r}}{r},
\label{eq:OPEP}
\end{equation}
where 
\begin{equation}
\alpha_\pi \equiv \frac{g_A^2 m_\pi^2}{16 \pi f_\pi^2}.
\end{equation}

Strictly speaking at second order in the EFT we should calculate pionic
corrections to this potential as well as including derivatives of
the regulated contact interaction. However, as a first attempt we 
include only these derivative terms, using as our second-order EFT 
interaction a sum of the interaction of Eqs.~(\ref{eq:VR}) and
(\ref{eq:sqwell}) and the OPEP (\ref{eq:OPEP}).

The radial Schr\"odinger equation in the $l = 0$ partial wave is then:
\begin{equation}
\left( -\frac{1}{M} - 2 C_2 \delta_{R}(r) \right) \frac{d^2
U}{dr^2} - 2 C_2 \left( \frac{dU}{dr} - \frac{U(r)}{r} \right)
\frac{d \delta_R}{dr} + \left( V_{\pi}(r) + C_0 \delta_{R}(r) - C_2
\frac{d^2 \delta_{R}}{dr^2} \right) U(r) = E U(r).
\label{eq:radialpion}
\end{equation}
The boundary condition at $r = R$ is identical to the case with only a
square well given in Eq.~(\ref{eq:bound}).

The equation (\ref{eq:radialpion}) cannot be solved analytically.
Therefore, we must solve the differential equation numerically and
impose a matching condition far from the origin in order to find the
phase shifts.  Once this is done, the analysis proceeds exactly as in
Section~\ref{sec:sqwell}.  We note that because the pion potential is
not compactly supported, it is no longer clear that the results of
Ref.~\cite{CP96} demand a lower bound on $R$.  We will, nevertheless,
see in Section~\ref{sec-lowerbd} that such a lower bound does exist in
this case.

\begin{figure}
\HideDisplacementBoxes
\centerline{\BoxedEPSF{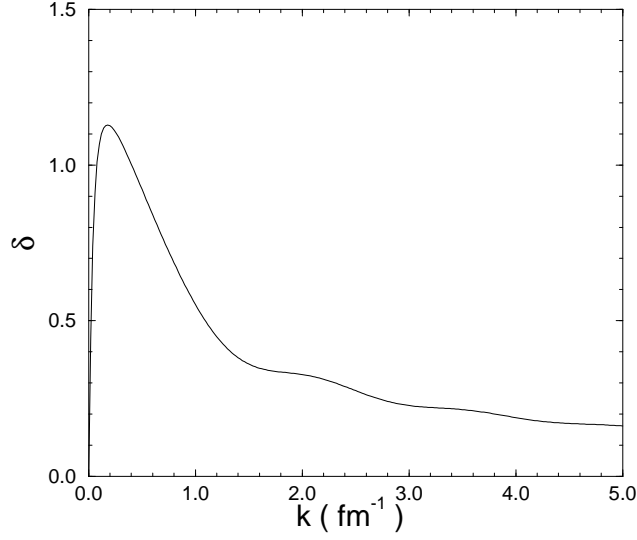 scaled 500}}
\caption{A plot of phase shifts in radians versus momentum for the
	pion potential plus an optimal well.
	These are calculated from the numerical solution of
	Eq.~(\ref{eq:radialpion}) with $R = 2.3\ {\rm fm}, C_0 =
	-3.3\ {\rm fm}^2$ and $C_2 = 0$.}	\label{f:optpion} 
\end{figure}

When we set $C_2 = 0$ and solve for the values of $R$ and $C_0$ which
reproduce the $\mbox{}^1 S_0$ NN scattering length and effective
range, we obtain an optimal well width of 2.3 fm with $C_0
= - 3.3\ {\rm fm}^2$.  A plot of the phase shifts as a function of $k$
for this potential is shown in Figure~\ref{f:optpion}. 

\begin{figure}
\HideDisplacementBoxes
\centerline{\BoxedEPSF{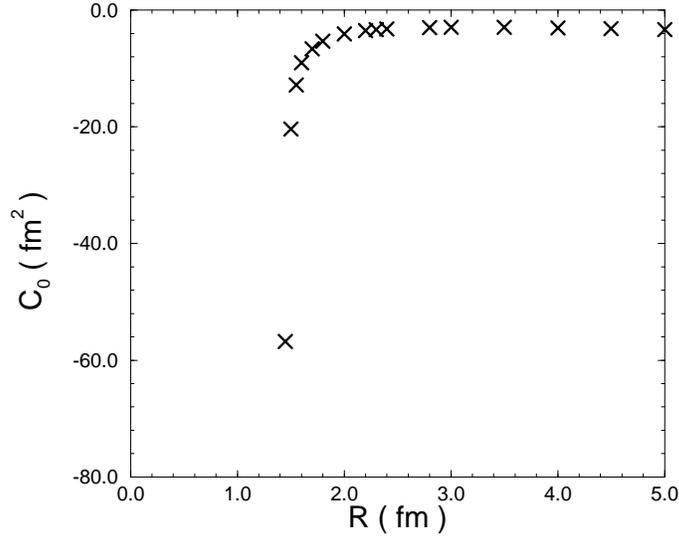 scaled 500}}
\caption{A plot of the renormalized values of the zeroth-order
	coefficient in the EFT expansion, $C_0$, from the
	numerical solution of Eq.~(\ref{eq:radialpion}) that reproduces the
	low-energy ${}^1S_0$ NN scattering data, versus the radius, $R$, 
	of the square well piece of the potential.} 
		\label{f:c0pion}
\end{figure}

\begin{figure}
\HideDisplacementBoxes
\centerline{\BoxedEPSF{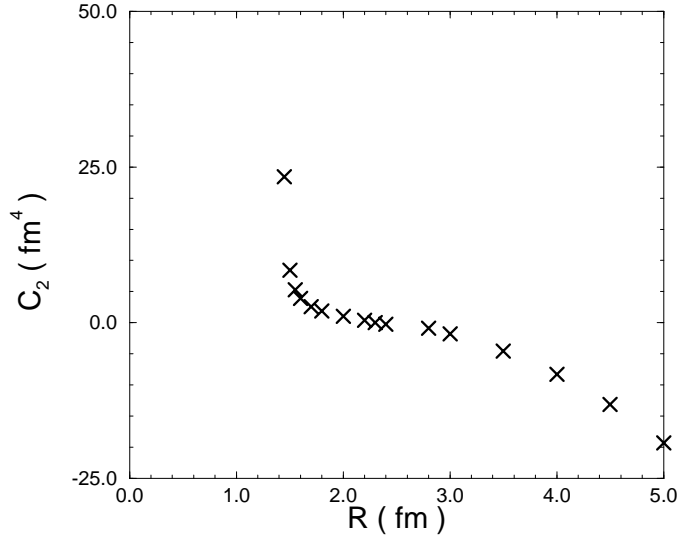 scaled 500}}
\caption{A plot of the renormalized values of the second-order
	coefficient in the EFT expansion, $C_2$, versus the radius,
	$R$, of the square well piece of the potential.}
		\label{f:c2pion}
\end{figure}

Next we renormalize the coefficients $C_0$ and $C_2$ for
other values of $R$. Figures~\ref{f:c0pion} and \ref{f:c2pion} plot the
renormalized values.  We observe that both $C_0$ and $C_2$ diverge
near $R=1.4 \mbox{ fm}$, indicating that this is the smallest well that can
be used to parameterize the NN interaction in the ${}^1S_0$ channel
when a one-pion exchange potential is included.

\begin{figure}
\HideDisplacementBoxes
\centerline{\BoxedEPSF{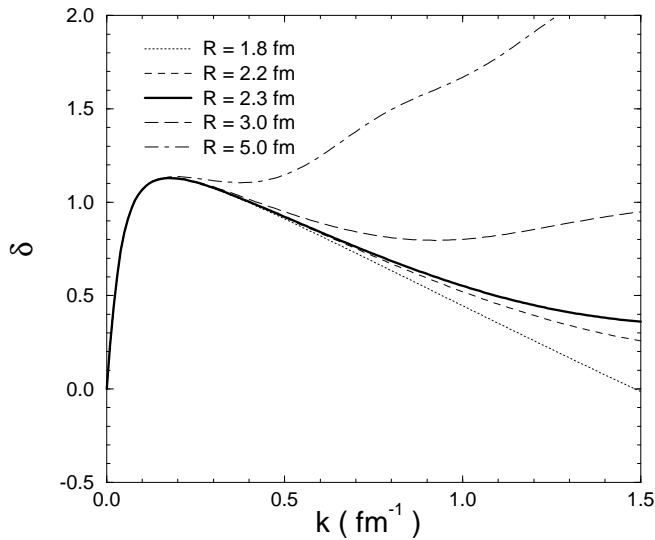 scaled 500}}
\caption{A plot of the phase shifts in radians versus momentum for
	various radii from the numerical solution of
	Eq.~(\ref{eq:radialpion}).  The phase shifts for the optimal
	well are shown in bold.} 
		\label{f:deltpion}
\end{figure}

Now that we have renormalized $C_0$ and $C_2$, these values can be used
to calculate the phase shifts for our second-order EFT potential.
A comparison of the phase shifts for several values of
$R$ is shown in Figure~\ref{f:deltpion}.  We observe the same qualitative
behavior
as was seen in Figure~\ref{f:rall}.  The phase shifts for all radii
agree up to a point---as must be, since we have fit the first
two terms in the effective range expansion.  At this point, the results for 
potentials with radii far
from optimal diverge rapidly.  However, for values of $R$ reasonably
close to the optimal value, the agreement persists to a considerably
higher momentum. 

Nevertheless, one sees that sensitivity to the particular regulator
chosen creeps into the observables at a momentum of $k \approx 0.5
\mbox{ fm}^{-1}$ (earlier if the well is ridiculously large). This is
not surprising since this is roughly the momentum at which details of
the well structure begin to be probed. One might think that this is an
argument for removing the regulator from the problem by taking $R$ to
zero. However, in this work we have shown that $R$ {\it cannot} be
decreased below about $1.4$ fm. Thus, we appear to be forced to use a
regulator which limits the applicability of the EFT to momenta below
some maximum, $\overline{\Lambda}$. This maximum lies considerably
below the expected range of validity of the EFT.  This is not a
desirable situation.  Similar
comments apply to the results of Fig.~\ref{f:rall}, but there the pion
is integrated out, and so $\overline{\Lambda}$ is of roughly the same
magnitude as the expected maximum momentum for the EFT.  What we see
in Fig.~\ref{f:deltpion} is that including the pion explicitly does not greatly
decrease the sensitivity of the results to the radius of the
short-range potential.  This occurs because the large effective range
of this problem means that the ``short-range'' potential cannot really
be made short range at all, but instead must extend out to at least
$1.4$ fm. This suggests the possibility of a breakdown in the scale separation
which is essential for power-counting arguments to be applicable.

\begin{figure}
\HideDisplacementBoxes
\centerline{\BoxedEPSF{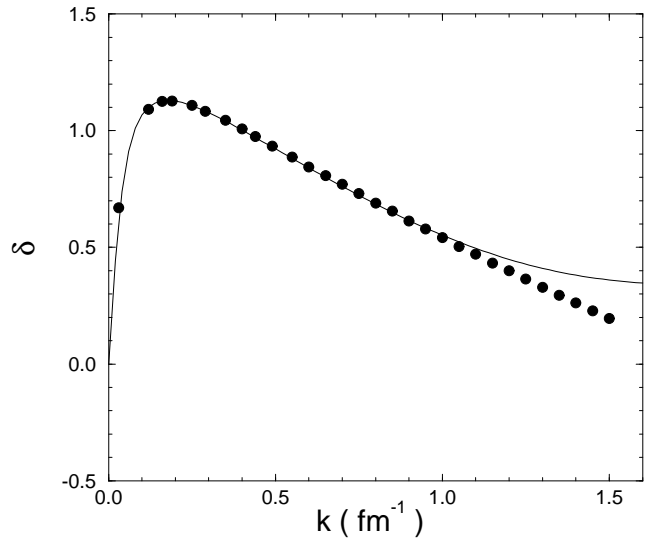 scaled 500}}
\caption{The solid line in this plot is the phase shift as a function
	of momentum for the OPE potential and an optimal square well.
	It is compared with data points from the Nijmegen $np$ partial
	wave analysis in the $\mbox{}^1 S_0$ channel~\protect\cite{St93}}
		\label{f:data}
\end{figure}

Figure~\ref{f:data} compares the phase shifts for the OPE potential
plus optimal well with the Nijmegen $np$ partial wave analysis for the
$\mbox{}^1 S_0$ channel~\cite{St93}.  This plot shows good agreement
up to the pion production threshold.  The divergence at higher momenta
is sharp but not unexpected since calculated phase shifts are positive
for all momenta while the experimental phase shift becomes negative
just above pion threshold. This behavior could be reproduced if we
tuned the potential specifically to do so.

However, we observe that there is really no reason to compare the
Nijmegen phase shifts with the results from the optimal well. The
phase shifts from any of the other wells whose phase shifts are
plotted in Fig.~\ref{f:deltpion} can be compared to the data too. Upon
doing this it is clear that the optimal well happens to be a potential
whose structure affects the phase shifts in a way which brings them
closer to to the experimental data. In other words, the good agreement
seen in Fig.~\ref{f:data} is fortuitous, and is dependent upon details of
the particular regulator used. We cannot expect such good agreement in
general.

\begin{figure}
\HideDisplacementBoxes
\centerline{\BoxedEPSF{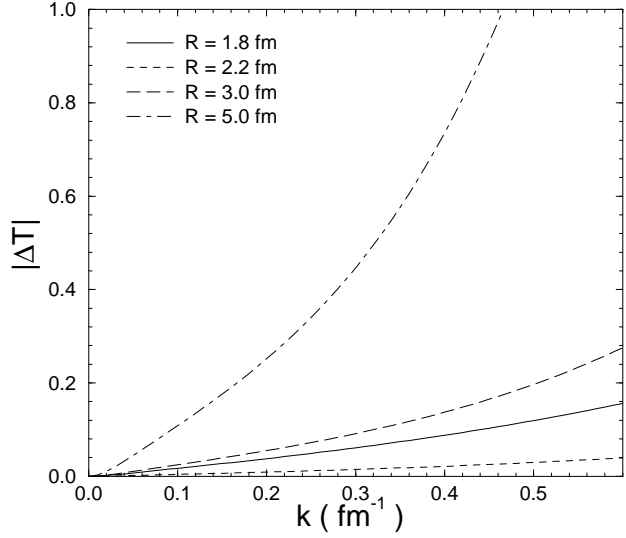 scaled 500}}
\caption{A plot of the fractional difference between the zeroth- and
	second-order t-matrices versus momentum for several values of
	$R$. For an optimal well, with $R$=2.3 fm, this quantity is
	identically zero.}
		\label{f:DeltaTpion}
\end{figure}

Finally, we once again compare the the zeroth- and second-order
t-matrices for different values of $R$.  We will examine the
fractional difference $\Delta T_{R}(k)$ as defined in
Eq.~(\ref{eq:DeltaT}).  The second-order t-matrix, $T^{(2)}_{R}(k)$,
is calculated using the pion potential and a square well with second
derivative terms. As in Section~\ref{sec:sqwell} the zeroth-order
t-matrix is calculated with only the zeroth-order potential
(\ref{eq:EFT0}), fitting the scattering length to match that of the NN
interaction in the ${}^1S_0$ channel.  The results are plotted in
Figure~\ref{f:DeltaTpion}.  Again the behavior is very similar to the
case where the pion was integrated out.  The inclusion of the pion has
not decreased the size of this quantity significantly, because the
``short-range'' square well is itself of roughly one-pion range.

\section{A lower-bound on the range of the short-range interaction}

\label{sec-lowerbd}

In the previous section we saw that the radius of the square well
potential could not be reduced below 1.4 fm if the scattering length
and effective range generated by the sum of a short-range and one-pion
exchange potential were to agree with the experimental results. In
fact, there exists a lower bound on this radius which is independent
of the choice of regulating potential. In this section we extend the
arguments of Ref.~\cite{CP96} in order to show that, if the theory is
to fit the scattering length and effective range, then there is a general
absolute lower bound on the range of the non-one-pion-exchange piece
of the interaction.

Consider the radial Schr\"odinger equation for S-wave scattering in
the case where both a one-pion exchange potential and an arbitrary
(possibly non-local) short-range potential, $V_R$ are present:
\begin{equation}
-\frac{1}{M}\frac{d^2u_E(r)}{dr^2} + \int \, dr' r V_R(r,r') r' u_E(r')
 - \alpha_{\pi} \frac{e^{-m_{\pi} r}}{r} u_E(r)=E u_E(r),
\label{eq:VRSE}
\end{equation}
with $u_E(0)=0$. Here $V_R(r,r')=0$ for $r > R$ or $r'>R$.
This, of course, was the case discussed in the previous
section for certain specific $V_R$s. 
However, in this section, instead
of enquiring as to the exact nature of $V_R$, we now consider a solution,
$v_E(r)$ of the equation
\begin{equation}
-\frac{1}{M}\frac{d^2v_E(r)}{dr^2}
 - \alpha_{\pi} \frac{e^{-m_{\pi} r}}{r} v_E(r)=E v_E(r).
\label{eq:noVR}
\end{equation}
$v_E$ is chosen so as to match onto the asymptotic wave function,
$u_E(r)=\sin(kr + \delta_E)$, with $k=\sqrt{ME}$ and $\delta_E$ the
experimental phase shift, as $r \rightarrow \infty$, and is normalized
so that $v_E(0)=1$. Suppose now that $\tilde{u}_E(r)$ is a solution of
(\ref{eq:VRSE}), with the parameters of $V_R$ adjusted so that
$\tilde{u}_E$ has the experimentally observed asymptotic
behavior. Suppose also that $\tilde{u}_E$ is normalized so that it
agrees with $v_E(r)$ at $r=R$. Given this normalization the two wave
functions agree on $[R,\infty)$.  However, they differ on $[0,R]$, in
that $\tilde{u}_E(0)=0$, while $v_E(0)=1$.

Now, going through the arguments displayed in \cite{CP96} yields:
\begin{equation}
\left. \frac{dv_2}{dr} \right|_{r=0} - \left. \frac{dv_1}{dr} \right|_{r=0}
= (k_2^2 - k_1^2) \int_0^\infty dr \, [v_2(r) v_1(r) - \tilde{u}_2(r) 
\tilde{u}_1(r)],
\label{eq:v2v1}
\end{equation}
where $v_2$ and $v_1$ are obtained by solving (\ref{eq:noVR}), with
the appropriate boundary conditions at $r=R$, at two different
energies $E_2$ and $E_1$. Similarly, $\tilde{u}_2$ and $\tilde{u}_1$
are solutions to (\ref{eq:VRSE}) at the same two energies. From
Eq.~(\ref{eq:v2v1}) we find
\begin{equation}
\frac{d}{dE} \left(\left. \frac{dv_E}{dr} \right|_{r=0}\right)
=M \int_0^\infty \, dr \, [v_E^2(r) - \tilde{u}_E^2(r)].
\end{equation}
Since $u_E(r)$ may be chosen to be real, and $u_E$ and $v_E$ 
agree for $r \geq R$, it follows that
\begin{equation}
\frac{d}{dE} \left(\left. \frac{dv_E}{dr} \right|_{r=0}\right)
\leq M \int_0^R dr \, v_E^2(r).
\label{eq:constraint}
\end{equation}

Once the asymptotic behavior of $v_E$ is specified the function
$v_E(r)$ is independent of $R$, depending only on the experimental
phase shift $\delta_E$. By contrast, the right-hand side is a function
of $R$, but only through the integral's
upper bound. Thus, if $\frac{d}{dE} \left(\left. \frac{dv_E}{dr}
\right|_{r=0}\right)$ is positive, then as $R$ is decreased towards zero 
a value of $R$ will be reached for which (\ref{eq:constraint}) will be
violated.  

In the case where there is no one-pion exchange interaction
the wavefunction $v_E(r)$ is
\begin{equation}
v_E(r)=\frac{\sin(kr + \delta_E)}{\sin(\delta_E)},
\end{equation}
and so, Eq.~(\ref{eq:constraint}) becomes:
\begin{equation}
\frac{d}{dE} (k \cot \delta_E) \leq 0,
\end{equation}
as claimed in the Introduction. As mentioned in Section~\ref{sec:sqwell}, 
this may be derived from an old result of Wigner~\cite{Wi55}. 

\begin{figure}
\HideDisplacementBoxes
\centerline{\BoxedEPSF{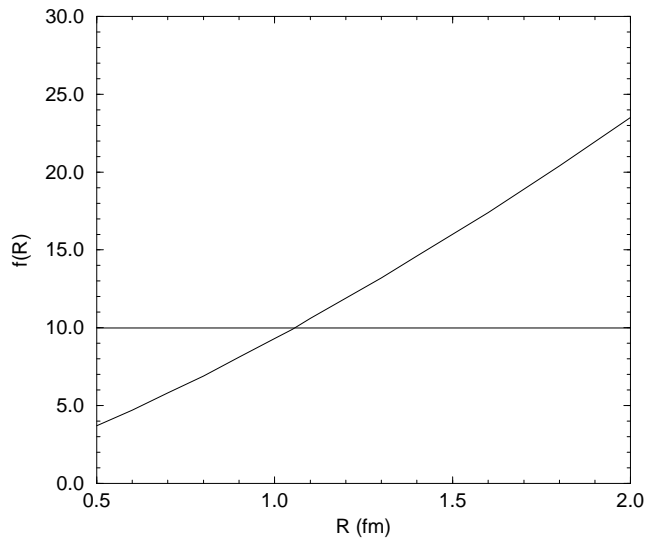 scaled 500}}
\caption{A comparison of the two sides of Eq.~(\ref{eq:constraint}),
	 evaluated at $E=0$ for NN scattering in the ${}^1S_0$ channel.
	 Observe that the function $f(R)$ is less than the
	 numerical bound (\ref{eq:numericalbd}) for radii less than
	 1.1 fm.}
\label{f:bound}
\end{figure}

In the case where one-pion exchange is included the function $v_E(r)$
may be calculated numerically given experimental phase shift data.
When this is done using the experimental values for $a$ and $r_e$
given in Eq.~(\ref{eq:exptalvals}) the result:
\begin{equation}
\frac{d}{dE} \left(\left. \frac{dv_E}{dr} \right|_{r=0}\right)_{E=0}
=10.0,
\label{eq:numericalbd}
\end{equation}
accurate to two significant figures, is obtained. (In practice this 
is best done by fixing the logarithmic derivative at some small, but finite,
distance, integrating out to large distances and matching to the 
experimental phase shifts, and only then integrating in to the origin.)
We may now check
whether Eq.~(\ref{eq:constraint}) is violated at $E=0$. In
Fig.~\ref{f:bound} we plot the right-hand side of
Eq.~(\ref{eq:constraint}), taken at $E=0$,
\begin{equation}
f(R) \equiv M \int_0^R dr \, v_0^2(r),
\label{eq:fdefn}
\end{equation}
as a function of $R$. The constraint (\ref{eq:constraint})
is violated once $R < 1.1$ fm. It follows that {\it any} short-range
potential of range less than 1.1 fm which is used in the Schr\"odinger
equation (\ref{eq:VRSE}) will not be able to fit the experimental
scattering length and effective range.

\section{Conclusions and Discussion}

In this paper we have shown how a finite-range potential can
parameterize the short-range physics in an effective field theory (EFT)
approach to the nucleon-nucleon interaction.  We have chosen a square-well 
potential since it allows a simple and clear analysis using elementary
quantum mechanics.  Other forms of potentials could be used 
and should have little effect on the results.  
We have considered both the case where
all exchanged particles are integrated out, and the case where a
one-pion exchange potential is retained.

We have done EFT calculations at both zeroth and second order.  In
doing so we have adopted the approach of Weinberg, and done the power
counting in our EFT in the potential, rather than in the amplitude.
The zeroth-order calculation then involves a potential which is the
sum of a square well and one-pion exchange (if explicit pions are
present), while our second-order EFT calculation adds derivatives of
the regulated contact interaction. (It should be noted that our
``second-order'' EFT calculation in the theory with pions ignores some
two-pion exchange graphs which are, strictly speaking, of the same EFT
order as graphs we have included in our potential. However, we believe
the inclusion of these graphs in the calculation will not
qualitatively alter our conclusions.)  For a given regulator parameter
the coefficients in the short-range interaction are fitted to the
${}^1S_0$ scattering length and effective range.

An optimal choice for the well radius was found in both the theory
with explicit pions and that without. At that well radius the
second-order piece of the potential is identically zero and yet the
${}^1S_0$ scattering length and effective range are reproduced.  When
one attempts to renormalize the coefficients in the EFT expansion so
as to fit the experimental data at other well radii, there turns out
to be a lower bound on the well radii for which this can be done
successfully.  This suggests that this method of regulating the theory
does not allow one to reach the limit of truly zero-range
interactions.  However, it should be noted that other, non-equivalent,
regulation schemes may allow one to define a Schr\"odinger equation
containing contact interactions.

For potentials with radii close to the optimal value the second-order
corrections to the t-matrix are small, and the phase shifts are
similar to those produced by the optimal well. This lack of
sensitivity to the regulator parameter indicates that it might be
possible to develop a systematic power-counting scheme along the lines
of the calculations performed in this paper.  However, our results
show that such a power-counting scheme can only hope to be successful
for momenta up to about 100--150 MeV. This is a range of validity which
excludes many interesting nuclear physics applications.  It is
somewhat surprising that the cutoff cannot be reduced
beyond 1.4 fm for the case of a square-well regulator.  Indeed, the
results of Section~\ref{sec-lowerbd} show that, no matter what
regulator is used, it must extend beyond $r=1.1 \mbox{ fm}$. This means that
the physics which, together with one-pion exchange, explains the
effective range in the ${}^1S_0$ channel, is not particularly 
short-ranged.

In the case where the OPE potential was included, we found
that the optimal well fits the experimental phase shifts well.
However, if other regulator parameters are chosen the data 
is not fit particularly successfully beyond about $0.5 \mbox{ fm}^{-1}$.

There are a number of points to observe in comparing the approach of
this paper with that recently advocated by Kaplan {\it et
al.}~\cite{Ka96}.  First, we note that the regulation of the delta
function interactions used in that work was completely different to
that employed here.  Whereas we have defined a contact interaction as
the limit of a sequence of square wells of decreasing radii, Kaplan
{\it et al.} have calculated sets of loop diagrams using true contact
interactions and then renormalized the resulting infinities using
dimensional regularization and the $\overline{{\rm MS}}$ scheme.  The
two approaches to regulating the delta function potential are not
equivalent. Specifically, in our approach the range of the short-range
potential cannot be taken to zero if the renormalization conditions on
its coefficients are to be satisfied.

Second, upon keeping the short-range potential of finite range, we see
that using the Weinberg approach to power counting allows the
explanation of both the scattering length and the effective range in
an EFT which is valid to momenta $k \approx 0.5 \mbox{ fm}^{-1}$.  This
is in contrast to the results of Kaplan {\it et al.} who found that,
given their method of regulation, if the scattering length and
effective range were to be explained in a Weinberg power-counting
approach then the domain of validity of the resulting EFT was very
small. It was this that led Kaplan {\it et al.} to define a new
power-counting scheme in which power counting was employed for the
inverse scattering amplitude. With the form of regulation used in this
paper power counting may always be applied to the potential without a
poor radius of convergence for the EFT resulting.  Of course, our
effective field theory expansion is open to question since our
regulator was never, and indeed can never be, removed from the theory.
However, we have demonstrated that when the regulator parameter is
kept finite and within sensible bounds, there is not great sensitivity
to it for processes involving momenta up to about 100 MeV.  A formal
power-counting scheme for an EFT approach where the range of the
short-range interaction is always kept finite remains to be worked
out.

Finally, from the standpoint of correctly reproducing the experimental
data, in the case without pion exchange the approach of
Ref.~\cite{Ka96} (which there is identical to effective range theory)
does considerably better than our ``optimal well''.  When the pion is
included, the two approaches appear to yield nearly identical results,
although the lack of an {\it a priori} reason for choosing the optimal
well over wells of other radii means that we cannot unambiguously say
that the phase shifts predicted by our approach are those shown in
Fig.~\ref{f:data}. Above 100 MeV the predictions from potentials in
which different regulator parameters were used differ considerably.
This is a troubling result, and casts doubt on the efficacy of the
approach discussed here. However, strictly speaking, the momenta to
which the EFT of Kaplan {\it et al.} applies is of roughly this size,
although a comparison of their results with experimental data
indicates that good agreement is obtained to much higher momentum than
100 MeV.  Thus, it remains to be seen if either approach can be
successfully used in describing other nuclear physics problems.

\acknowledgements{We thank Bira van Kolck for a number of useful
conversations, and for helpful comments on the manuscript. We also
thank Silas Beane for his comments on the manuscript. We are grateful
to the U.~S. Department of Energy for its support under grant
no. DE-FG02-93ER-40762.}

\end{document}